\documentclass{emulateapj}

\newcommand{\etal}{et al.}
\def\gsim{\lower 2pt \hbox{$\, \buildrel {\scriptstyle >}\over
{\scriptstyle \sim}\,$}}
\def\lsim{\lower 2pt \hbox{$\, \buildrel {\scriptstyle <}\over
{\scriptstyle \sim}\,$}}

\def\cvi{C~{\scriptsize VI}}
\def\cii{C~{\scriptsize II}}
\def\ciii{C~{\scriptsize III}}
\def\civ{C~{\scriptsize IV}}
\def\nvii{N~{\scriptsize VII}}
\def\nv{N~{\scriptsize V}}

\def\oviii{O~{\scriptsize VIII}}
\def\ovii{O~{\scriptsize VII}}
\def\ovi{O~{\scriptsize VI}}
\def\oi{O~{\scriptsize I}}
\def\oii{O~{\scriptsize II}}

\def\neviii{Ne~{\scriptsize VIII}}
\def\neix{Ne~{\scriptsize IX}}

\def\HI{H~{\scriptsize I}}

\def\siiv{Si~{\scriptsize IV}}
\def\siiii{Si~{\scriptsize III}}
\def\siii{Si~{\scriptsize II}}
\def\feii{Fe~{\scriptsize II}}

\shortauthors{Yao \etal}
\shorttitle{X-raying the intergalactic \ovi\ absorbers}

\slugcomment{\em Accepted for publication in the  Astrophysical Journal}
\begin{document}

\title{X-raying the intergalactic \ovi\ absorbers}
\author{Y. Yao\altaffilmark{1}, T. M. Tripp\altaffilmark{2}, 
	Q. D. Wang\altaffilmark{2}, C. W. Danforth\altaffilmark{1}, 
	C. R. Canizares\altaffilmark{3}, \\
	J. M. Shull\altaffilmark{1},
	H. L. Marshall\altaffilmark{3},
	and L. Song\altaffilmark{2}
}
\altaffiltext{1}{Center for Astrophysics and Space Astronomy, University of
Colorado, 389 UCB, Boulder, CO 80309; yangsen.yao, michael.shull, charles.danforth@colorado.edu}
\altaffiltext{2}{Department of Astronomy, University of Massachusetts, 
  Amherst, MA 01003; tripp, wqd, limin@astro.umass.edu}
\altaffiltext{3}{Massachusetts Institute of Technology (MIT) Kavli Institute 
 for Astrophysics and Space Research, 70 Vassar Street, Cambridge, MA 02139;
	crc, hermanm@space.mit.edu}

\begin{abstract}

The observed intergalactic \ovi\ absorbers at $z>0$ have been regarded
as a significant reservoir of the ``missing baryons''. However, to
fully understand how these absorbers contribute to the baryon
inventory, it is crucial to determine whether the systems are
collisionally ionized or photoionized (or both). Using the identified
intergalactic \ovi\ absorbers as tracers, we search for the
corresponding X-ray absorption lines, which are useful for finding the
missing baryons and for revealing the nature of the \ovi\
absorbers. Stacking the {\sl Chandra} grating spectra along six AGN
sight lines, we obtain three spectra with signal-to-noise ratios of
32, 28, and 10 per 12.5 m\AA\ spectral bin around the expected
\ovii~K$\alpha$ wavelength.  These spectra correspond to \ovi\
absorbers with various dynamic properties.  We find no detectable
\neix, \ovii, \oviii, \nvii, or \cvi\ absorption lines in the spectra,
but the high counting statistics allows us to obtain firm upper limits
on the corresponding ionic column densities (in particular $N_{\rm
OVII}\lsim10~N_{\rm OVI}$ on average at the 95\% confidence
level). Jointly analyzing these non-detected X-ray lines with the
averaged \ovi\ column density, we further limit the average
temperature of the \ovi-bearing gas to be $T\lsim10^{5.7}$ K in
collisional ionization equilibrium.  We discuss the implications of
these results for physical properties of the putative warm-hot
intergalactic medium and its detection in future X-ray observations.

\end{abstract}


\section{Introduction }
\label{sec:intro}

Hydrodynamic simulations of cosmic large-scale structure indicate
that, in the local universe ($z\lsim1$), the shock-heated gas in the
tenuous warm-hot intergalactic medium (WHIM) at temperatures
$T\sim10^{5-7}$ K contains a substantial amount of baryonic matter
\citep{cen99, dav01}, providing a possible solution for the so-called
``missing baryons'' problem (e.g., \citealt{per92, fuk98}). In a
collisionally ionized gas at these temperatures, the most abundant
heavy elements (e.g., C, N, O, and Ne) are in their high ionization
(e.g., Li-, He-, and H-like) states \citep{sut93}, whose K- and
L-shell transitions are in the X-ray and far-ultraviolet (far-UV)
wavelength bands, respectively.  The existence of the WHIM has been
suggested through detections of X-ray emission from high-density
regions near galaxies, groups, and clusters (e.g., \citealt{wang97,
wang04, fin03, sol07, man07}; see also \citealt{dur08} for a review), 
but these regions are expected to
contain only a small portion of the WHIM gas. The majority is located
in low-density cosmic-web filaments \citep{dav01}, whose emission is
hard to detect with current X-ray and far-UV telescopes.  These
filaments are most easily probed through absorption lines imprinted on
spectra of background sources.

With the high sensitivity spectrographs aboard the {\sl Hubble Space
Telescope} (HST) and the {\sl Far Ultraviolet Spectroscopic Explorer}
(FUSE), the intergalactic \ovi\ absorption lines (at rest-frame
wavelengths $1031.93$ \AA\ and 1037.62 \AA) have routinely been
detected in many background AGN spectra (e.g., \citealt{sav98, tri00a,
tri00, shu03, pro04, dan05, coo08}). However, the nature of these
\ovi\ absorbers is still under debate. In a survey of the
intergalactic medium (IGM) absorption lines, \citet{dan08} found a
good correlation in column density and a similar power-law slope of
the column density distribution $d{\cal{N}}/dz$ of \ovi\ and \nv,
which are distinct from those of \HI, \ciii, and \siiii.  These
features, along with the theoretical expectations of high post-shock
temperatures and long cooling times of the WHIM, motivated them to
argue for a multiphase nature of the IGM. In this picture, \ovi\ and
\nv\ trace the canonical WHIM at temperatures of $10^{5-6}$ K, while
the low ionization ions like \cii-\civ\ and \siii-\siiv\ are
predominantly photoionized. But in two other independent surveys that
include a number of common sight lines to those used by \citet{dan08},
\citet{tri08} and \citet{tho08} found that 30-40\% of the \ovi\
absorbers have velocity centroids that are well aligned with those of
the associated \HI\ absorbers and that some absorber temperatures
inferred from the line widths, $T < 10^5$ K, are well below that
expected in the canonical WHIM. \citet{tri08} also showed that these
velocity-aligned absorbers can be naturally explained by
photoionization models. These latter findings are consistent with
previous investigations along individual sight lines based on the
kinematic and chemical properties of the high- and low-ionization
absorbers (e.g., \citealt{tri00, sav02, sem04, pro04, leh06}).
However, the UV data usually do not have high enough quality to
uniquely constrain the ionization mechanism; \citet{tri08} also showed
that the \ovi\ lines could alternatively arise in hot interface layers
on the surface of low-ionization clouds if the \ovi\ systems are
always multiphase media. Current data cannot rule out this
possibility, and many \ovi\ absorbers are fully consistent with this
hypothesis (see \S~4.2 in \citealt{tri08}). So far, the strongest
indication of the collisional ionization origin of the \ovi\ absorbers
is the detection of \neviii\ absorption lines (at rest-frame
wavelengths 770.4 \AA\ and 780.3 \AA) toward HE~0226-4110 at $z_{\rm
a}=0.2070$ and toward 3C~263 at $z_{\rm a}=0.3257$ \citep{sav05,
nar08}. However, a systematic search toward other sight lines has
failed to find similar systems \citep{leh06}.

The claimed detections of the WHIM in the X-ray band are still
controversial.  In the spectrum of PKS~2155-304 observed with the {\sl
Chandra} Advanced CCD Imaging Spectrometer (ACIS), \citet{fang02}
reported an \oviii\ K$\alpha$ absorption line at $z=0.0553$, claiming
the first detection of the WHIM in the X-ray. This detection has not
been confirmed, although it cannot be ruled out either, by the {\sl
Chandra} High Resolution Camera (HRC) observations and the subsequent
{\sl XMM-Newton} Reflection Grating Spectrometer (RGS) observations
\citep{cag04, wil07, fang07}.  \citet{nic05} reported detections of
two $z>0$ \ovii\ WHIM systems in the {\sl Chandra} spectra of Mrk~421,
which were not confirmed from the {\sl XMM-Newton} RGS observations of
the same source \citep{kaa06, ras07}.  \citet{kaa06} also questioned
the statistical significance of the Mrk~421 detections in the {\sl
Chandra} spectra. The other claimed $z>0$ X-ray WHIM absorptions are
mostly detected at marginal significance (e.g., \citealt{mat03}).
Very recently, \citet{buote09} have reported a $3\sigma$ detection of
\ovii\ K$\alpha$ absorption affiliated with a large-scale structure
(the Sculptor Wall). The \ovii\ and \oviii\ absorption lines at
$z\simeq0$, which may be partially due to the WHIM in the Local Group
(e.g., \citealt{nic02, fang03}), have been relatively well detected,
but these absorptions are severely confused by the contributions from
the Galactic hot gas close to (or within) the Milky Way disk
\citep{sem03, yao05, yao07, wang05, fang06, bre07, yao08}.

With the controversial X-ray detections of the WHIM and the debatable
nature of the observed far-UV absorbers, two key questions still
remain open: (1) Does the WHIM exist? (2) If it does, what are its
physical properties?

The highly ionized X-ray absorption lines are believed to be useful
for finding the WHIM gas and for probing the nature of the \ovi\
absorbers.  The \ovii\ ion can trace gas over a broad temperature
range, and its column density is expected to be $\gsim10$ times higher
than that of \ovi\ in a shock-heated gas with shock temperatures
$T\gsim10^{5.7}$ K (for overdensity $\delta\sim10-100$; e.g.,
\citealt{fur05}).  Moreover, cosmological simulations predict that hot
\ovi\ absorbers should have comparably strong (or even stronger)
affiliated \ovii\ absorption lines (see, e.g., Figures 8-10 in
\citealt{cen06} and Figure 14 in \citealt{chen03}). However, because
of the limited sensitivity and spectral resolution of the current
X-ray observatories like {\sl Chandra} and {\sl XMM-Newton}, searching
for X-ray lines in the WHIM is currently a difficult task.  For
instance, the column densities of the X-ray absorbing ions, even for
the most promising one (\ovii ), are expected to be small with $N_{\rm
OVII}\approx 10^{15}~\rm{cm^{-2}}$ (e.g., \citealt{chen03}). For the
most part, the background AGN are also relatively faint in the X-ray
band.  Consequently, very long exposures (see \S~\ref{sec:dis}) are
required to collect enough photons to conduct a blind search for X-ray
absorption lines from the WHIM along a random line of sight. These
difficulties contribute to the controversies and frustrations
surrounding the reported X-ray detections.

In this work, we search for the X-ray absorption lines of \ovii, as
well as \oviii, \neix, \nvii, and \cvi, by extensively exploring the
archived {\sl Chandra} grating observations.  We use the identified
\ovi\ absorbers as tracers to avoid uncertainties of a blind search.
We also develop a technique to stack all the applicable X-ray
observations and obtain spectra with unprecedented counting
statistics.

The paper is organized as follows. In \S~\ref{sec:data}, we list the 
identified \ovi\ absorbers and the corresponding {\sl Chandra} observations 
utilized in this work, and describe the data reduction process.
In \S~\ref{sec:method}, we present our search method and stack the X-ray
observations in the rest frame of the observed \ovi\ absorbers.
We search for the X-ray absorption lines in the stacked spectra,
present the results of the data analysis in \S~\ref{sec:results},
and discuss the implications of our results in \S~\ref{sec:dis}.

\section{\ovi\ absorbers, {\sl Chandra} observations, and data reduction}
\label{sec:data}

There are $\sim80$ identified IGM \ovi\ systems at redshifts of
$0<z_{\rm abs}<0.5$ toward $\sim30$ AGN sight lines \citep{sav02,
sem04, dan06, dan05, dan08, tri08, tho08}.  {\sl Chandra} has observed
12 of these AGN with high spectral resolution grating instruments. To
avoid the potential confusion caused by the absorbers intrinsic to the
AGN in identifying the IGM \ovi\ systems and in searching for X-ray
lines, we do not use the five AGN (Mrk~279, NGC~5548, Mrk~509,
Ark~564, and NGC~7469) toward which warm absorbers have been detected
in the X-ray band (e.g., \citealt{sco04, sco05, kaa02, yaq03, mat04}).
{\sl Chandra} observations of the five AGN could contribute an
additional $<10\%$ in total to the spectral counts of our final
spectra (see below), so excluding them in the data analysis does not
significantly affect our results.  To avoid the confusion caused by
possible absorber misidentification or noise features, we only used
the \ovi\ systems that were detected at $>3\sigma$ significance
levels.  We further excluded the \emph{proximate} absorbers that are
within $\sim2,000~{\rm km~s^{-1}}$ to the AGN redshifts; \citet{tri08}
have shown that the majority of the proximate \ovi\ systems are found
within this velocity interval at low redshifts. The proximate systems
can be high-velocity AGN ejecta and/or can be located near the central
engine of the AGN. Consequently, these systems should be excluded when
searching for WHIM absorbers.  

Table~\ref{tab:ovi} summarizes our \ovi\ absorber sample, including
the redshifts of the \ovi\ systems ($z_{\rm OVI}$), $\lambda$1031.93
equivalent widths, dispersion velocities ($b$), and column densities
($N_{\rm OVI}$) along the six sight lines used in this work.  For
those absorbers at very close redshifts along a single sight line that
cannot be distinguished with the {\sl Chandra} resolution, only the
systems with higher EWs were used. Galaxy surveys at low $z$ show that
most of these \ovi\ absorbers are usually found within several hundred
kpc of intervening galaxies along the sight lines (\citealt{tri00,
sav02, shu03, sem04, tum05, pro06, sto06, tri06, coo08}).  The
investigation of the connection between galaxy environment and X-ray
absorbers is underway and will be presented in a separate paper. In
this work, we focus on searching for corresponding X-ray absorption of
the \ovi-bearing gas.

\begin{deluxetable*}{lclcccl}%[t]
\tablewidth{0pt}
\tabletypesize{\footnotesize}
\tablecaption{\small Properties of the identified \ovi\ absorbers 
\label{tab:ovi}}
\tablehead{
                 &               &		     & EW     & $b$ &  \\
Source Name      & $z_{\rm AGN}$ &  $z_{\rm OVI}$ & (m\AA) & (km~s$^{-1}$) & $\log[N_{\rm OVI}(\rm cm^{-2})]$ & Class$^b$}
\startdata
H1821+643        & 0.2970	 & 0.0244            & $27\pm8$  & $21^{+9}_{-6}$   & $13.44\pm0.1$  & Sm \\
                 &      	 & 0.1214 	     & $97\pm14$ & $76^{+13}_{-11}$ & $13.97\pm0.06$ & Cm, St\\ 
		 &		 & 0.2133            & $39\pm9$  & $28\pm5$         & $13.54\pm0.06$ & Cm  \\
		 &		 & 0.2250$^a$        & $190\pm10$& $45\pm2$         & $14.27\pm0.02$ & Cm, St\\
                 &		 & 0.2264$^a$        & $32\pm4$  & $16\pm2$         & $13.51\pm0.04$ & Cm \\ 
                 &		 & 0.2453            & $51\pm7$  & $26\pm2$         & $13.71\pm0.03$ & Sm \\ 
                 &		 & 0.2666            & $45\pm7$  & $25\pm3$         & $13.63\pm0.04$ & Sm  \\
\hline
3C 273           & 0.1583	 & 0.0033  	     & $31\pm7$  & $51^{+11}_{-9}$  & $13.44\pm0.07$ & Cm\\ 
                 &		 & 0.0902            & $16\pm3$  & $22\pm6$         & $13.18\pm0.06$ & Cm\\ 
                 &		 & 0.1200            & $24\pm3$  & $8\pm3$          & $13.37\pm0.04$ & Sm \\
\hline 
PG 1116+215      & 0.1765	 & 0.0593            & $63\pm9$   & $36^{+11}_{-8}$ & $13.52\pm0.08$ & Sm \\ 
		 &    		 & 0.1385 	     & $83\pm16$  & $36\pm7$        & $13.97\pm0.06$ & Cm, St\\
		 &    		 & 0.1655 	     & $111\pm9$  & $32^{+21}_{-13}$& $14.08\pm0.04$ & Cm, St\\
\hline
PKS 2155-304	 & 0.1165	 & 0.0540$^a$        & $32\pm5$  & $14\pm6$   & $13.63\pm0.12$ & Cm \\
		 &   		 & 0.0572$^a$ 	     & $44\pm11$ & $24\pm7$   & $13.57\pm0.09$ & Cm \\ 
\hline
Ton S180         & 0.0620	 & 0.0456            & $62\pm14$& 20    & $13.74\pm0.06$ & Sm \\ 
\hline
PG 1211+143      & 0.0809	 & 0.0511            & 187	& $53^{+10}_{-9}$ & $14.21\pm0.08$ & Cm, St \\
      		 & 	 	 & 0.0645$^a$ 	     & $144\pm32$& $54\pm11$   & $14.16\pm0.08$ & Cm, St \\ 
		 &		 & 0.0649$^a$        & $55\pm10$& 21   & $13.83\pm0.11$  & Cm \\ 
\hline
averaged$^c$     &		 &             & 45.4      & 28   & 13.62 & All\\ 
averaged$^d$     &		 &             & 49.0      & 32   & 13.65 & Cm \\
averaged$^e$     &		 &             & 131.7     & 59   & 14.11 & St 
\enddata
\tablecomments{
Errors are listed in 1$\sigma$ range.
The values of the first three sources (H1821+643, 3C~273, and PG1116+215) 
are adopted from \citet{tri08}, and the remaining values are adopted from
\citet{dan08} except for those at the redshift 0.0511 along
the PG~1211+143 sight line, which are adopted from \citet{tum05}.
$^a$ Toward an individual sight line, these absorbers cannot be 
distinguished with {\sl Chandra} resolution, and the system with higher 
equivalent width of \ovi\ is used.
$^b$ Classification of the \ovi\ absorbers; ``Cm'', ``Sm'', and ``St''
denote complex, simple, and strong \ovi\ systems, respectively.
$^{c,d,e}$ Values are weight averaged with respect to the spectral counts at the \ovii\
K$\alpha$ wavelength (21.602 \AA) in the shifted spectra for all, complex, 
and strong \ovi\ absorbers, respectively (\S~\ref{sec:results}).}
\end{deluxetable*}

\begin{figure*}%[h]
\plotone{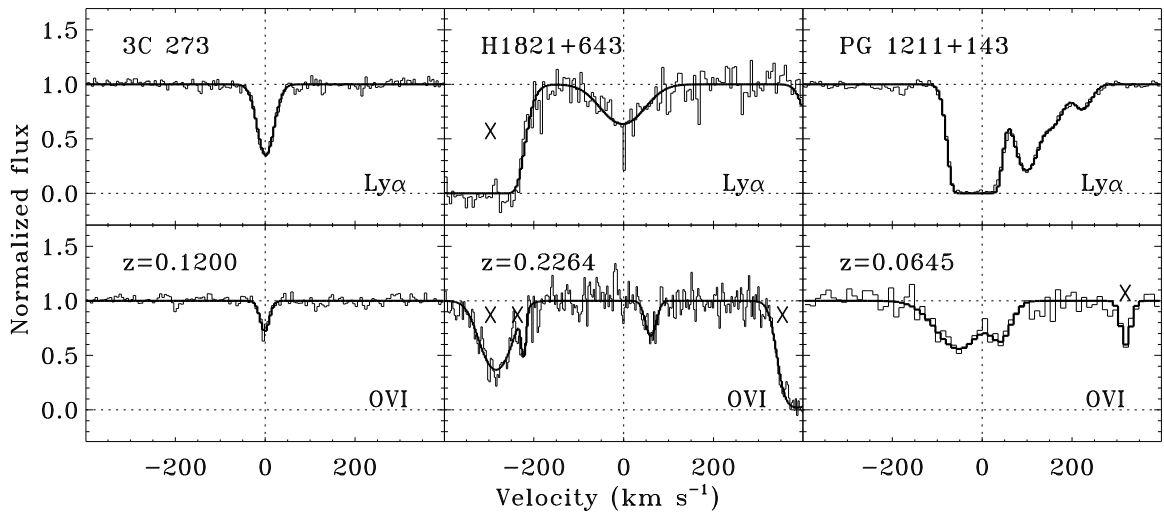}
\caption{From left to right,
	examples of the simple, complex, and strong  \ovi\ 
	absorbers and the corresponding Ly$\alpha$ systems, along 
	the 3C~273, H1821+643, and PG~1211+143 sight lines
	\citep{tri08, dan08}, respectively.
	Absorbers that are not associated with the labeled systems are
	marked with ``X''.
   \label{fig:lineclass} }
\end{figure*}

We follow the criteria used in \citet{tri08} to classify the \ovi\
absorbers along the six AGN sight lines into \emph{simple} and
\emph{complex} systems. In simple systems, the velocity centroids of
\HI\ and \ovi\ are well aligned (within $2\sigma$ of their velocity
uncertainties).  In complex absorbers, these velocity centroids are
well separated ($>2\sigma$ apart) and/or there are multiple low- and
high-ionization stages (e.g., \cii-\civ, \siii-\siiii, or \nv, in
addition to \ovi) that indicate the presence of multiple ionization
phases.  We also classify the \ovi\ absorbers with equivalent width
(EW) $>80$ m\AA\ as \emph{strong} systems (e.g., \citealt{fang01}).
\citet{tri08} classified the \ovi\ absorbers along three of the six
sight lines. Using these criteria, we classified all the \ovi\
absorbers reported by \citet{dan06} and \citet{dan08} toward the other
three sight lines.  Figure~\ref{fig:lineclass} presents examples of
each classified system.

As of 2008 May 6, there were $\sim100$ archived {\sl Chandra} grating
observations of the six selected targets. In this work, we did not use
non-standard observations with uncertain calibrations.  We also
excluded several short observations (with an individual exposure
$\lsim10$ ks) of PKS~2155-304 that contribute $<5\%$ to the total
spectral counts of the source.  Table~\ref{tab:obs} summarizes the
number of observations and the total exposure used in this work.

\begin{deluxetable}{lccr}
\tablewidth{0pt}
\tablecaption{\small {\sl Chandra} grating observations
  of the selected targets \label{tab:obs}}
\tablehead{
                 &	     & \#. obs.          & Exp.\\
Src. Name        & $z_{\rm AGN}$ & (LETG)  & (ks)}
\startdata
H1821+643        & 0.297     & 5           &600  \\
3C 273           & 0.158     &16          &360  \\
PG 1116+215      & 0.176     & 1           &89   \\
PKS 2155-304     & 0.117     &36          &760 \\
Ton S180         & 0.062     & 1           &80   \\
PG 1211+143      & 0.081     & 3           &141  \\
\hline
\multicolumn{2}{l}{sub total:}& 62          &~~2030 
\enddata
\end{deluxetable}

For ACIS grating observations, we followed the procedures described in
\citet{yao05, yao07} to calibrate the data, extract spectra, and
calculate corresponding response functions (RSPs). The same energy
grid was applied throughout these procedures.  We utilized the medium
energy grating (MEG) spectra of the High Energy Transmission Grating
(HETG; \citealt{can05}) observations and only used the first grating
order spectra of all observations.

For each HRC grating observation, we followed the steps presented in
\citet{wang05} to obtain the first order RSP (FRSP) and
order-overlapped RSP (ORSP) and spectra. We then extracted the first
order spectra by subtracting the difference of the best-fit-model
predicted channel counts between the ORSP and the FRSP from the
order-overlapped spectra.

\begin{figure*}%[h]
\plotone{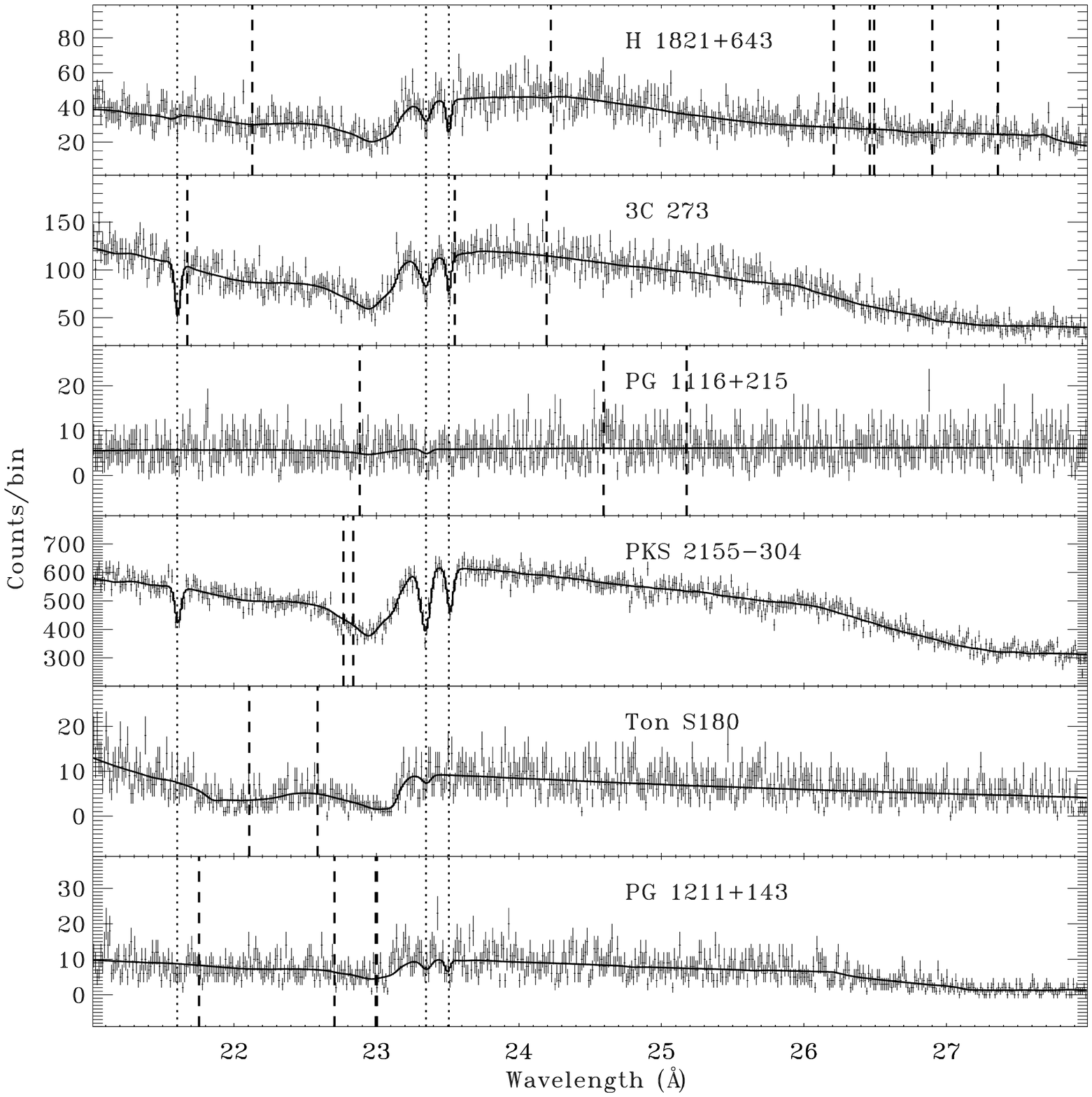}
\caption{{\sl Chandra} spectra of the six sources around the \ovii\ K$\alpha$ 
   line and the best-fit continuua.
   The vertical dotted lines mark the Galactic \ovii, \oi, and \oii\
   K$\alpha$ lines at 21.602, 23.508, and 23.348 \AA, respectively.
   The thick dashed lines mark the expected positions of the intergalactic 
   \ovii\ K$\alpha$ lines at the corresponding $z_{\rm OVI}$
   (Table~\ref{tab:ovi}).
   The bin size is 12.5 m\AA.
   \label{fig:src} }
\end{figure*}

Spectra from positive and negative grating arms of each observation
were then co-added, and multiple observations toward a single sight line
were further stacked to enhance the counting statistics.
Visual inspection revealed no significant and 
consistent K$\alpha$ lines of \ovii, \oviii, \neix, \nvii, or \cvi\ at the 
corresponding redshift $z_{\rm OVI}$ along individual sight lines. 
Figure~\ref{fig:src} shows parts of the final spectra of the six
AGN around the expected IGM \ovii\ K$\alpha$ lines. 

\begin{figure*}%[h]
\plotone{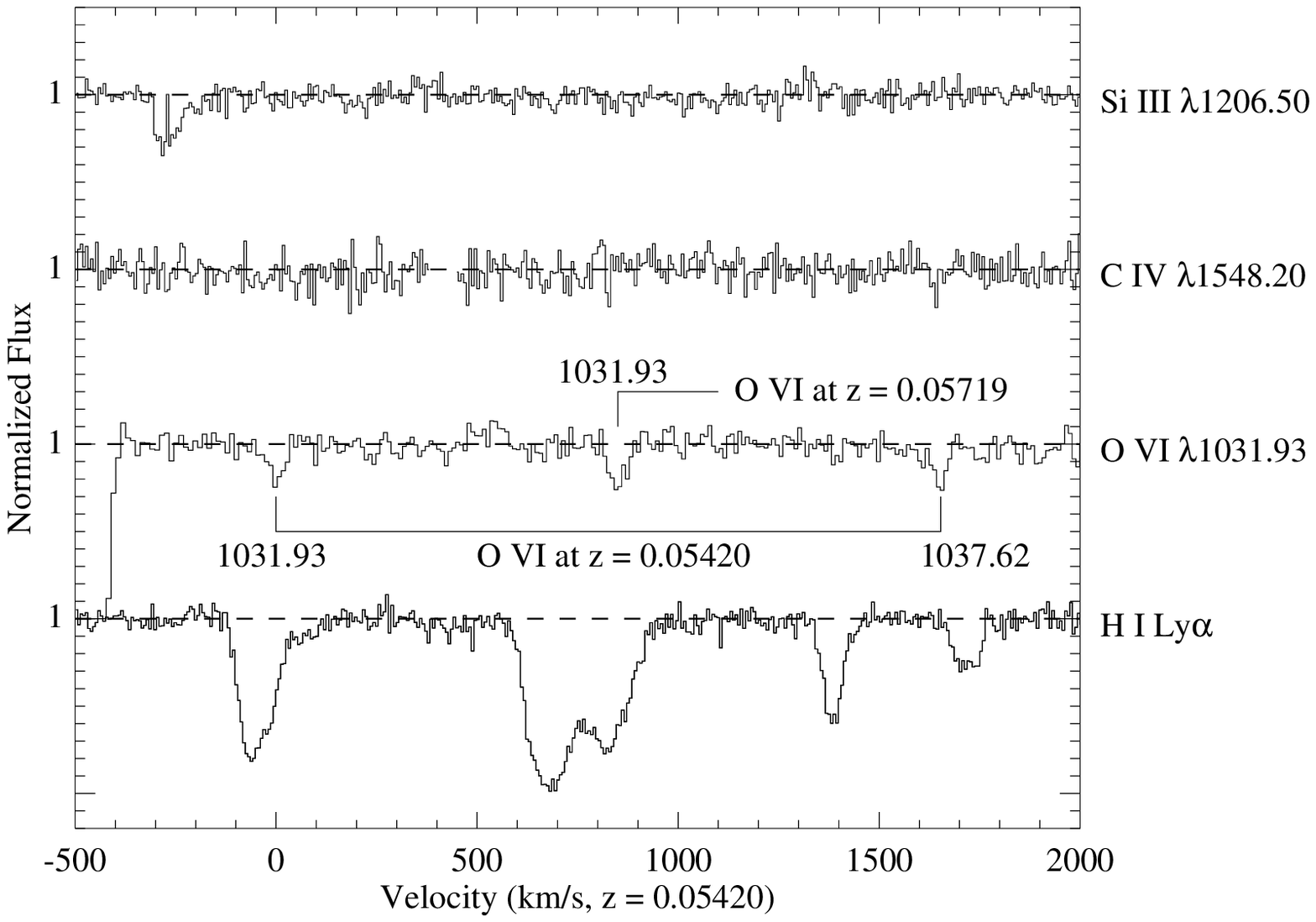}
\caption{Metal absorption lines associated with a cluster 
   of \HI\ Ly$\alpha$ lines detected in the PKS~2155-304 
   spectrum, plotted vs. velocity with $v=0~{\rm km~s^{-1}}$ 
   at $z_{\rm abs}=0.0542$. Both lines of the \ovi\ doublet 
   are clearly detected at $z_{\rm abs}=0.0542$ and are 
   marked. Only the \ovi\ $\lambda1031.93$ transition is
   detected at $z_{\rm abs}=0.0572$, but the weaker \ovi\ 
   $\lambda1037.62$ line is blended with a strong Galactic 
   \feii\ transition. 
   \label{fig:pks-UV} }
\end{figure*}

Spectra along the PKS~2155--304 sight line deserve to be mentioned 
particularly because of their excellent spectral quality 
(Fig.~\ref{fig:src}). Toward this sight line, \citet{shu03} detected two 
\ovi\ absorption systems at redshifts $z=0.054$ and $0.057$, associated with
a cluster of seven \HI\ absorbers and a group of galaxies \citep{shu98}.
Figure~\ref{fig:pks-UV} shows the detected \ovi\ and other
non-detected metal absorption lines along with the cluster of
\HI\ lines. On the {\sl Chandra} ACIS-grating spectrum,
the claimed $z=0.0553$ \oviii\ line at $\lambda_{\rm obs}=20.02$ 
\AA\ is clearly visible with an EW of 5.9(3.1, 8.6) m\AA\ (90\% confidence 
range), which is consistent with that reported by \citet{fang02, fang07}. 
There is no other X-ray line at this redshift. The discrepancy 
between the HRC-grating, the {\sl XMM-Newton} RGS (with an accumulated exposure time
of 980 ks), and the ACIS-grating spectra still exists; the former two spectra
yield upper limits of $EW<3.2$ m\AA\ and $<3.7$ m\AA, respectively
(see also \S~\ref{sec:intro} and references therein).
The line in the co-added HRC+ACIS spectrum is severely diluted
with $EW<4.2$ m\AA.

\section{An effective stacking and searching strategy}
\label{sec:method}

To facilitate an effective search for X-ray IGM absorption lines, we
first blueshifted the X-ray spectral data by the observed \ovi\
absorber redshifts $z_{\rm OVI}$ and reconstructed the corresponding
RSPs accordingly.  The data are originally in forms of detected counts
distributed in a wavelength grid that was chosen to be the same for
all the X-ray spectra.  The rest-frame wavelength $\lambda_0$ is
related to the observed wavelength $\lambda_{\rm obs}$ by
\begin{equation} \label{equ:shift}
\lambda_0 = \lambda_{\rm obs} /(1+z_{\rm OVI}). 
\end{equation}
For each absorber, we first use this relationship to re-calculate the grid 
boundaries of the corresponding X-ray spectrum and then cast the 
spectrum and the RSP onto a 
rest-frame spectral grid, which is chosen to be the original one. 
The re-calculated spectrum, both shifted and compressed, generally no longer 
matches this chosen rest-frame grid. Therefore, some of the re-calculated 
spectral bins fall completely within individual rest-frame grid intervals, 
whereas others enclose a grid boundary. In this latter case, we split the 
counts in such a bin into two parts and assign them to the two grid intervals 
adjacent to the enclosed boundary; the weights used in the splitting are 
proportional to the wavelength overlaps of the bin with the two intervals. In 
the same way, we re-calculate the wavelength boundaries and the distribution 
probability in the corresponding RSP file. We repeat this procedure for all 
the absorbers to obtain their respective rest-frame spectral and RSP files;
for individual sight lines with $n$ intervening \ovi\ systems 
(Table~\ref{tab:ovi}), we obtained $n$ shifted spectra and RSPs.

\begin{figure*}%[h]
\plotone{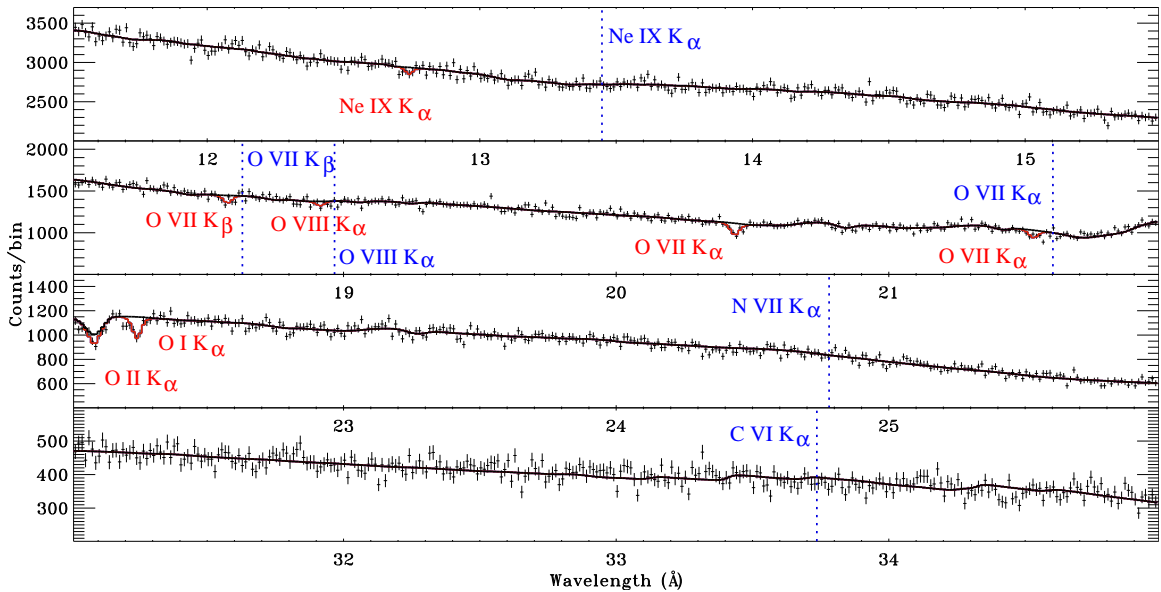}
\caption{Four ranges of the final stacked spectrum 
	that includes shifted spectra corresponding to all 
	16 \ovi\ systems (Table~\ref{tab:ovi}).
	The thick lines mark the best-fit continuum, the red lines 
	and text mark the blue-shifted Galactic absorptions 
	(mainly contributed 
	from PKS~2155--304 and 3C~273 sight lines), and blue vertical lines
	mark the positions of the expected intergalactic neon, 
	oxygen, nitrogen, and carbon absorption lines at the corresponding
	rest-frame wavelengths. The bin-size is 12.5 m\AA.
   \label{fig:all} }
\end{figure*}

\begin{figure*}%[h]
\plotone{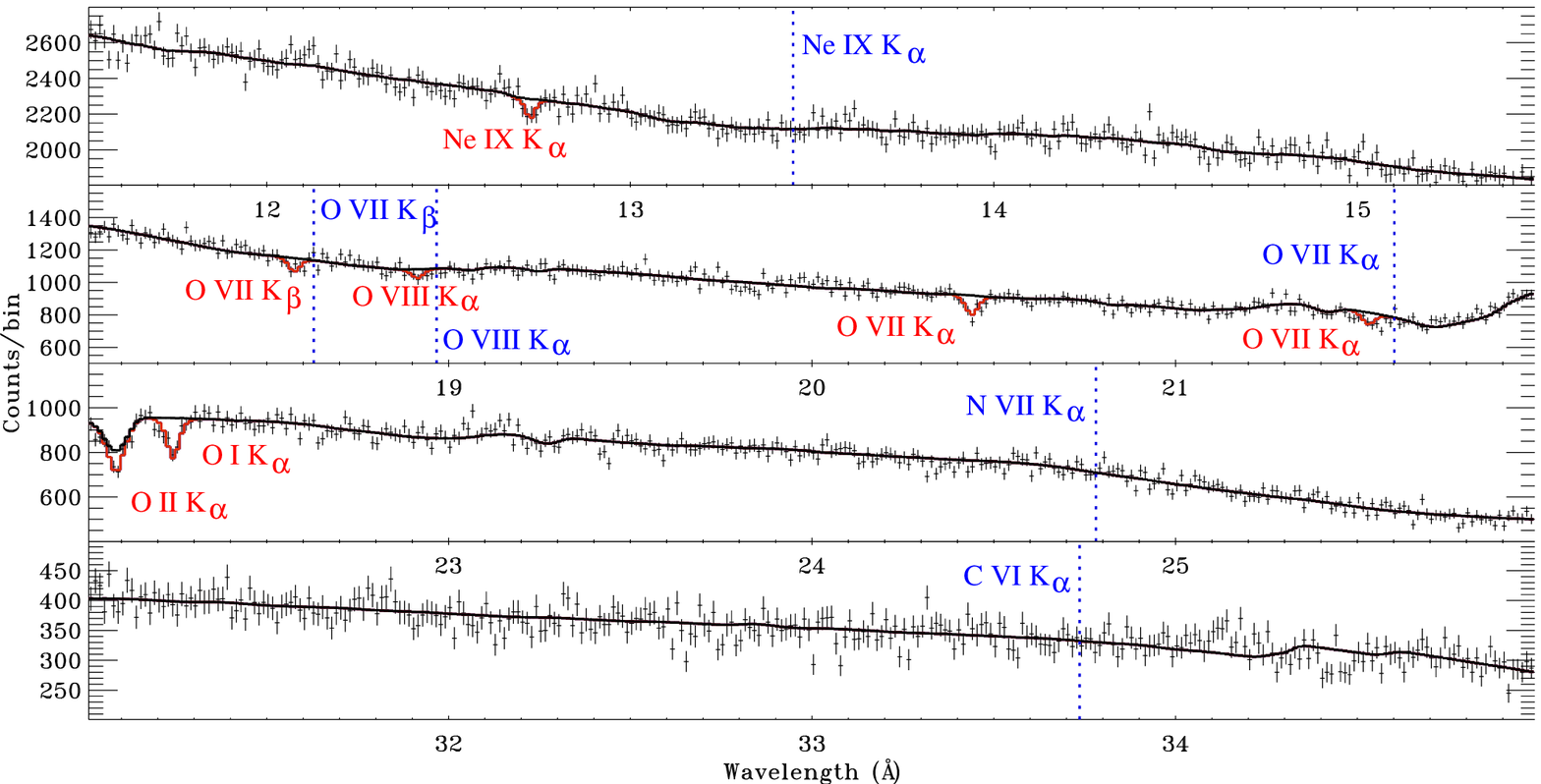}
\caption{Same as Figure~\ref{fig:all}, except that stacked spectrum 
	contains only spectra corresponding to the complex \ovi\ systems
	(Table~\ref{tab:ovi}).
   \label{fig:cm} }
\end{figure*}

\begin{figure*}%[h]
\plotone{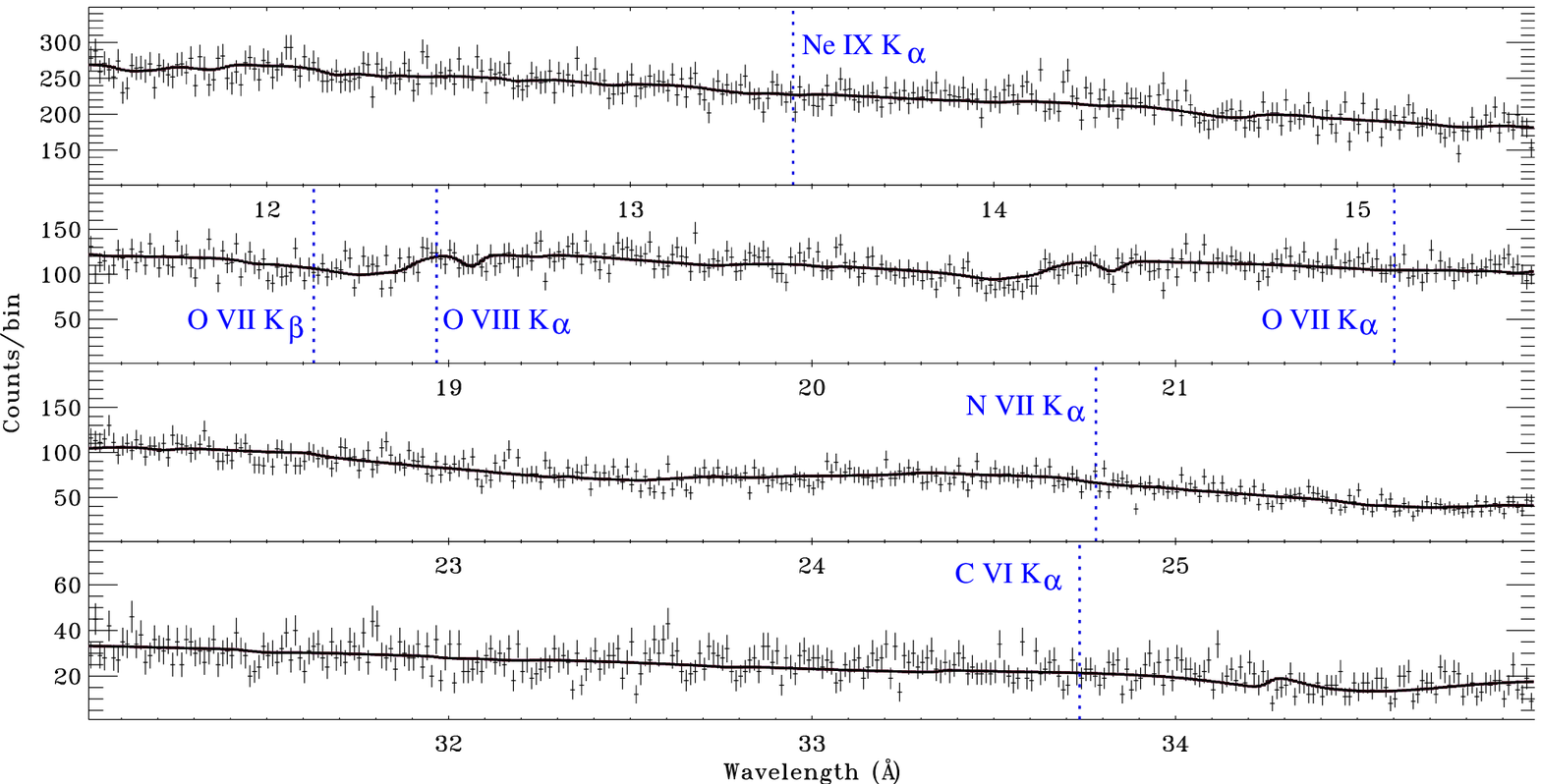}
\caption{Same as Figure~\ref{fig:all}, except that stacked spectrum
	contains only spectra corresponding to the strong \ovi\ systems
	(Table~\ref{tab:ovi}).
   \label{fig:st} }
\end{figure*}

To enhance the counting statistics, we co-added these shifted spectra
to form a single stacked spectrum and RSP. We constructed three
different stacked spectra based on the various \ovi\ absorber
classifications (\S \ref{sec:data}).  Figure~\ref{fig:all} illustrates
the stacked spectrum that contains {\it all} of the shifted spectra
corresponding to the 16 \ovi\ systems used in this work
(Table~\ref{tab:ovi}) regardless of their
classification. Figure~\ref{fig:cm} presents the stacked spectrum for
only the {\it complex} \ovi\ systems, and Figure~\ref{fig:st} is the
stacked spectrum for only the strong \ovi\ systems
(Table~\ref{tab:ovi}). To avoid any potential bias in our results
caused by the large contribution of the PKS~2155-304 sight line in the
stacked spectra (Fig.~\ref{fig:src}) and by its peculiarity in
detecting only the \oviii\ line (\S~\ref{sec:intro}), we also obtained
two similar stacked spectra and RSPs like those presented in
Figures~\ref{fig:all} and \ref{fig:cm}, but without the PKS~2155-304
contribution.  For the same reasons, we also report two sets of
results in the following sections when applicable, with and without
the contribution of PKS~2155-304.  It is interesting to point out
that, because of the poor spectral resolution of the X-ray
instruments, absorbers at very close redshifts/velocities along a
single sight line that cannot be distinguished with the {\sl Chandra}
LETG resolution ($\sim750~{\rm km~s^{-1}}$) are automatically
stacked/merged as one velocity component.

Because each spectrum has been shifted by the corresponding $z_{\rm
OVI}$ before being stacked, the X-ray IGM K$\alpha$ absorption lines
of \neix, \oviii, \ovii, \nvii, and \cvi, if they are associated with
the \ovi-bearing gas, are expected to be at rest-frame wavelengths of
13.448, 18.967, 21.602, 24.781, and 33.736 \AA, respectively, in the
stacked spectra.

\section{Analysis and Results}
\label{sec:results}

We now search for and measure the X-ray absorption lines in the
stacked spectra.  There are no extragalactic \ovii, \oviii, \neix,
\nvii, or \cvi\ absorption lines apparent in any of the stacked
spectra (Figs.~\ref{fig:all}--\ref{fig:st}).  Adding Gaussian profiles
at the corresponding rest-frame wavelengths, we obtain upper limits to
EWs of these lines. To obtain column densities of these ions, we first
calculate the mean dispersion velocity ($\overline{b}$) and column
density $\overline{N_{\rm OVI}}$ of different classified \ovi\ systems
(Table~\ref{tab:ovi}) by weighting them with respect to the spectral
counts at the wavelengths of the expected \ovii\ K$\alpha$ lines
(Fig.~\ref{fig:src}).  Replacing the Gaussian profiles with the
absorption line model {\sl absline}
\footnote{
In modeling a single absorption line, this model is similar to the
curve-of-growth analysis. But it can be used to jointly analyze 
multiple absorption lines at the same time. For a detailed description of the
model, see \citet{yao05}.}
\citep{yao05} and fixing the $b$ of different ions to 
$\overline{b}$, we then estimate the column density upper limits of 
these ions as reported in Table~\ref{tab:upper}.

\begin{deluxetable*}{cccccc}
\tablecaption{\small The 95\% confidence upper limits of K$\alpha$ absorption lines
	\label{tab:upper}}
\tablehead{
line              & \neix  & \oviii & \ovii  & \nvii  & \cvi \\
$\lambda$ (\AA)   & 13.448 & 18.967 & 21.602 & 24.781 & 33.736 }
\startdata
EW (m\AA)$^a$     & 1.0(1.5)      &  2.2(2.6)      & 1.1(2.6)       &  1.2(2.0)      & 3.6(8.1) \\ 
log[N(cm$^{-2}$)]$^a$ & 15.04(15.29)     &  15.51(15.57)    & 14.62(15.12)     &  14.85(15.06)    & 15.06(15.71) \\ 
\hline
EW (m\AA)$^b$     & 1.5(3.0)      &  2.3(3.0)      & 1.4(2.9)       &  1.3(3.0)      & 4.4(8.2) \\ 
log[N(cm$^{-2}$)]$^b$ & 15.28(15.71)  &  15.53(15.62)    & 14.75(15.15)     &  14.90(15.21)    & 15.15 (15.76)\\ 
\hline
EW (m\AA)$^c$     & 3.8      & 3.1       & 3.95       & 4.3       & 8.0 \\
log[N(cm$^{-2}$)]$^c$ & 15.75     &  15.63    & 15.19     &  15.33    & ~15.36 
\enddata
\tablecomments{Superscripts $a$, $b$, and $c$ denote the upper limits obtained 
	from stacked X-ray spectra corresponding to all, complex, and
	strong \ovi\ absorbers, respectively. Values in parenthesis indicate 
	limits constrained from the spectra without 
	contribution from the PKS~2155-304 sight line.
}
\end{deluxetable*}

With these upper limits to the ionic column densities and calculated
$\overline{N_{\rm OVI}}$, we can probe thermal properties of the
\ovi-bearing gas. For a gas in collisional ionization equilibrium
(CIE), the column density ratio between \ovii\ and \ovi\ provides a
diagnostic of gas temperature \citep{sut93, yao05}.  Assuming a single
gas temperature, we jointly analyze the non-detected \ovii\ with the
$\overline{N_{\rm OVI}}$ (Tables~\ref{tab:ovi} and \ref{tab:upper}),
and obtain a temperature upper limit of $\log[T{\rm (K)}]<5.7$ at
the 95\% confidence level from the three spectra corresponding to all,
complex, and strong \ovi\ absorbers.  Including the non-detected lines
of \neix, \nvii, and \cvi\ in the joint analysis (assuming relative
solar abundances of Ne/O, N/O and C/O; \citealt{and89}) does not
further constrain the temperature.  From the spectrum without the
contribution of PKS~2155--304, we obtain the same upper limit.  The
reason why these spectra yield nearly the same temperature upper limit
is the tightly constrained ratio of $N_{\rm OVII}/N_{\rm OVI}$ from
the various spectra from $\lsim10$ to $\lsim35$ (see
Tables~\ref{tab:ovi} and \ref{tab:upper}), compared to 8.9, 26.2, and
63.1 for $\log[T{\rm (K)}]=5.6$, 5.7, and 5.8, respectively, for gas
in CIE (see \citealt{sut93} and Fig.~6 in \citealt{tri01}).

\section{Discussion}
\label{sec:dis}

We have presented a search for X-ray absorption lines produced in the
WHIM by using the identified far-UV \ovi\ absorbers as tracers for
stacking the archived {\sl Chandra} observations. The three final
stacked spectra, corresponding to all, complex, and strong \ovi\
absorbers, have signal-to-noise ratios of $\sim32$, 28, and 10,
respectively, per 12.5 m\AA\ spectral bin around the \ovii\ K$\alpha$
wavelength.  There are no detectable X-ray absorption lines at the
expected wavelengths in these spectra.  We have obtained upper limits
to EWs of the K$\alpha$ lines of \ovii, \oviii, \neix, \nvii, and
\cvi\ and their column densities. Combining these non-detected lines
with the average \ovi\ column density, we have also derived an upper
limit to the temperature of the \ovi-bearing gas.
 
The X-ray measurements of the \ovi\ systems with different physical
properties in principle can reveal the ionization mechanism of the
intervening gas.  While most observed \ovi\ absorbers could arise in
interfaces between the low- and high-ionization phases (e.g.,
\citealt{dan08, tri08}), compared to the velocity well-aligned simple
\ovi\ absorbers, the dynamically complex \ovi\ systems provide more
direct evidence for the multiphase nature of the absorbers (e.g.,
\citealt{tri01, tri08, shu03, sem04, sav05}).  Cosmological
simulations of large-structure formation also predict that the \ovi\
systems with larger equivalent widths are more likely to be
collisionally ionized \citep{cen01, fang01}. If the \ovi\ absorbers
utilized in this work are indeed multiphase and trace the shock-heated
IGM, the hot \ovii-bearing gas is expected to be either surrounding
the cool absorbers (traced by the absorption lines of Ly$\alpha$,
\cii, \siii, etc.)  or itself containing the observed \ovi\ or a
combination of both. The column densities of \ovii\ (and/or \oviii)
are expected to be at least an order of magnitude more than that of
\ovi.  But the current X-ray observations only constrain the $N_{\rm
OVII}/N_{\rm OVI}\lsim10$ on average (Tables~\ref{tab:ovi} and
\ref{tab:upper}), which does not strongly favor the collisional
ionization scenario of the \ovi\ absorbers for shock temperatures
$T>10^{5.7}$ K \citep{fur05}.  However, these data still cannot rule
out the photoionization scenario, in which $N_{\rm OVII}/N_{\rm OVI}$
is expected to be $\lsim3$ (see Fig.~6 in \citealt{tri01} and
\citealt{fur05}).

There are several possibilities or their combinations
that may cause the non-detections of the X-ray lines:

1. Some of the \ovi\ systems may be photoionized or partly
photoionized.  The purely photoionized systems may not contribute to
the baryon inventory (\S~\ref{sec:intro}) since they could have
already been counted through Ly$\alpha$ absorbers.  Since the density
of the IGM is very low, photoionization could play an important role
in producing the highly ionized oxygen species (especially \ovi
\footnote{Higher ionization stages are harder to produce by
  photoionization, but we note that \citet{hel98} have argued
  that in the IGM, even \ovii\ and \oviii\ are predominantly 
  \emph{photoionized} by the X-ray background.}) in the WHIM besides the
  gravitational shocks (e.g., \citealt{chen03, tri08}). In this case,
  the $N_{\rm OVII}/N_{\rm OVI}$ is expected to be lower than that in
  a solely collisionally ionized gas.  Simulations show that, while
  the EWs of \ovi\ and \ovii\ are fairly well correlated in the
  putative WHIM in a broad overdensity range ($\delta\sim10-100$), for
  the \ovi-absorbers with $EW$(\ovi)$>34$ m\AA, there are $\lsim20\%$
  of the associated \ovii-absorbers with $EW$(\ovii)$>2$ m\AA\ (see
  Figs.~13-15 in \citealt{chen03} and Table~\ref{tab:upper}).

2. Thermal properties of the WHIM may vary among different systems and
sight lines. The \ovii\ and/or \oviii\ absorptions are expected to be
strong in some systems and weak in others, but the ``strong'' X-ray
absorption signals that originate from the shock-heated gas could have
been diluted in the stacked spectra.

3. The characteristic temperature of the hot gas may be too high
(e.g., $T>3\times10^6$ K, as observed in the intragroup and
intracluster medium) to produce observable \ovii\ or \oviii.  In this
case, the interface between hot and cool IGM could still be present to
be responsible for the observed \ovi\ absorbers, but the average
temperature upper limit ($\log[T{\rm (K)}]<5.7$) derived from the
ratio of $N_{\rm OVII}/N_{\rm OVI}$ (\S~\ref{sec:results}) only
applies to the vicinity of the \ovi-bearing gas. This may be the case
for the $z=0.0553$ absorber along the PKS 2155-304 sight line where
the \oviii\ line was detected but the \ovii\ was not
(\S~\ref{sec:intro} and references therein).

4. Lastly, the \ovi\ absorbers might only trace the warm part of the
WHIM at temperatures $T\lsim3\times10^5$ K, and the hot ($T\sim10^6$
K) IGM is not necessarily always co-located with the warm gas. If this
is the case, because of the small ionization fraction of \ovi\ at high
temperatures, the \ovi\ absorption lines produced in the hot gas might
not have been detected along most of the sight lines. The possible
large dispersion velocity of the hot gas could also make the \ovi\
absorption feature shallower and harder to detect.  The high
sensitivity of the Cosmic Origins Spectrograph, which is scheduled to
be installed on {\sl HST} in spring 2009, will provide a great
opportunity to search for such weak \ovi\ absorbers (at $z>0.12$) and
to examine this possibility.

The results obtained in this work also have important implications
regarding future observations of the WHIM with the current and the
future X-ray observatories.  With the still limited spectral
resolution [even in the proposed {\sl International X-ray observatory}
(IXO), $E/\Delta E\sim3,000$, which is still $\sim10-15$ times less
than those of current far-UV instruments], it will be challenging to
conduct a blind search for the X-ray absorption line ``forest''
produced in the WHIM. A more effective strategy is still to use the
identified \ovi\ and/or Ly$\alpha$ absorbers as references, as
implemented in this work (also see \citealt{chen03, dan05,
dan08}). However, if the six sight lines (Table~\ref{tab:ovi}) used in
this work fairly sample the WHIM filaments over the whole sky, the gas
only contains $\lsim10^{15}~{\rm cm^{-2}}$ of \ovii\ in column density
and the \ovii\ K$\alpha$ EW $\lsim2.5$ m\AA\ in absorption on average
(Table~\ref{tab:upper}). To detect a weak line with EW$\sim2$ m\AA\ in
an AGN spectrum with a flux of $4\times10^{-12}~{\rm
ergs~s^{-1}~cm^{-2}~keV^{-1}}$ around the line centroid, our 1,000-run
simulations for an exposure of 50 Ms with the {\sl Chandra} ACIS-LETG
only yield 375 detections at $EW/\Delta EW\gsim3\sigma$ significance
level. In comparison, for the {\sl IXO} with $E/\Delta E\sim3,000$ and
effective area of $\sim3,000~{\rm cm^2}$, our simulations indicate
that a 100 ks observation will have a 90\% probability of detecting
such a weak line at $\gsim3.5\sigma$ significance level.

\acknowledgements 

We are grateful to Renyue Cen for useful discussion on non-detection of 
X-ray absorption lines.
This work was made possible by {\it Chandra} archival research grant
AR7-8014.  Additional support for this research was provided by NASA
grant NNX08AC14G, provided to the University of Colorado to support
data analysis and scientific discoveries related to the Cosmic Origins
Spectrograph on the Hubble Space Telescope, and partly through the
Smithsonian Astrophysical Observatory contract SV3-73016 to MIT for
support of the Chandra X-Ray Center under contract NAS 08-03060.  TMT
and LS also acknowledge support for this work from NASA grant
NNX08AJ44G.

%\newpage

\end{document}